\documentclass{vgtc}                          

\ifpdf
  \pdfoutput=1\relax                   
  \usepackage{graphicx}                
  \DeclareGraphicsExtensions{.pdf,.png,.jpg,.jpeg} 
\else
  \usepackage{graphicx}                
  \DeclareGraphicsExtensions{.eps}     
\fi%

\graphicspath{{figures/}{pictures/}{images/}{./}} 

\usepackage{microtype}                 
\PassOptionsToPackage{warn}{textcomp}  
\usepackage{textcomp}                  
\usepackage{mathptmx}                  
\usepackage{times}                     

\usepackage{cite}                      
\usepackage{tabu}                      
\usepackage{booktabs}                  
\usepackage[square,sort,comma,numbers]{natbib}
\usepackage{multibib}

\newcites{Add}{Additional references}

\usepackage{hyperref}
\usepackage{url}

\onlineid{0}

\vgtccategory{Research}

\vgtcinsertpkg

\title{Review of Persuasive User Interface as Strategy for Technology Addiction in Virtual Environments}

 \author{Fachrina Dewi Puspitasari\thanks{e-mail: fachrina.puspitasari@kaist.ac.kr} %
 \and Lik-Hang Lee\thanks{e-mail:
 likhang.lee@kaist.ac.kr}} %
 \affiliation{\scriptsize Korea Advanced Institute of Science and Technology, Republic of Korea}

\abstract{
In the era of virtuality, the increasingly ubiquitous technology bears the challenge of excessive user dependency, also known as user addiction. Augmented reality (AR) and virtual reality (VR) have become increasingly integrated into daily life. Although discussions about the drawbacks of these technologies are abundant, their exploration for solutions is still rare. Thus, using the PRISMA methodology, this paper reviewed the literature on technology addiction and persuasive technology. After describing the key research trends, the paper summed up nine persuasive elements of user interfaces (UIs) that AR and VR developers could add to their apps to make them less addictive. Furthermore, this review paper encourages more research into a persuasive strategy for controlling user dependency in virtual-physical blended cyberspace.
} 
\keywords{Persuasive technology, Augmented reality, Virtual reality, Technology addiction, User interfaces, Metaverse.}

\CCScatlist{
  \CCScatTwelve{Human-centered computing}{Visualization}{Visualization design and evaluation methods}{}
}

\begin{document}




\firstsection{Introduction}
\maketitle
Excessive use of technology also raises various health problems, in addition to harming personal development. An example of disheartening news came in March 2021 from West Java province, Indonesia. A teenager died of brain damage, which was suspected to be caused by  
excessive exposure to mobile phone radiation~\citeAdd{pradana_2021}. The case of teenagers suffering from mobile phone addiction has become increasingly common. For example, in 2020, 98 teenagers became outpatients in the mental rehabilitation unit in the same province~\citeAdd{pradana_2021}. In the following two months, another fourteen teenagers underwent the same case~\citeAdd{pradana_2021}.

As technology becomes increasingly ubiquitous, controlling its usage is becoming difficult. Ubiquity creates the illusion that always indulging in technology is common. Moreover, in the metaverse era where everything is virtual, humans cannot stay away from technology. Thus, this raises the phenomenon of addiction to technology.

While research on the drawbacks of the virtual environment is numerous, their solutions are still insufficiently addressed. Based on this scarcity, this review paper aims to address the potential strategies to alleviate addiction in the virtual environment, particularly in augmented reality (AR) and virtual reality (VR). The techniques integrate the approach of persuasive technology to develop an unobtrusive and incremental influence on the addicts. Thus, aside from complementing the literature on addiction in the virtual environment, this review paper will also enrich the exploration of literature in persuasive technology that today is still heavily centered on wellbeing. Practically, strategies to be discussed later will potentially serve as 
a guideline for AR/VR developers to create engaging yet responsible applications.

\begin{figure}
{
  \centering
  \includegraphics[width=0.995\linewidth]{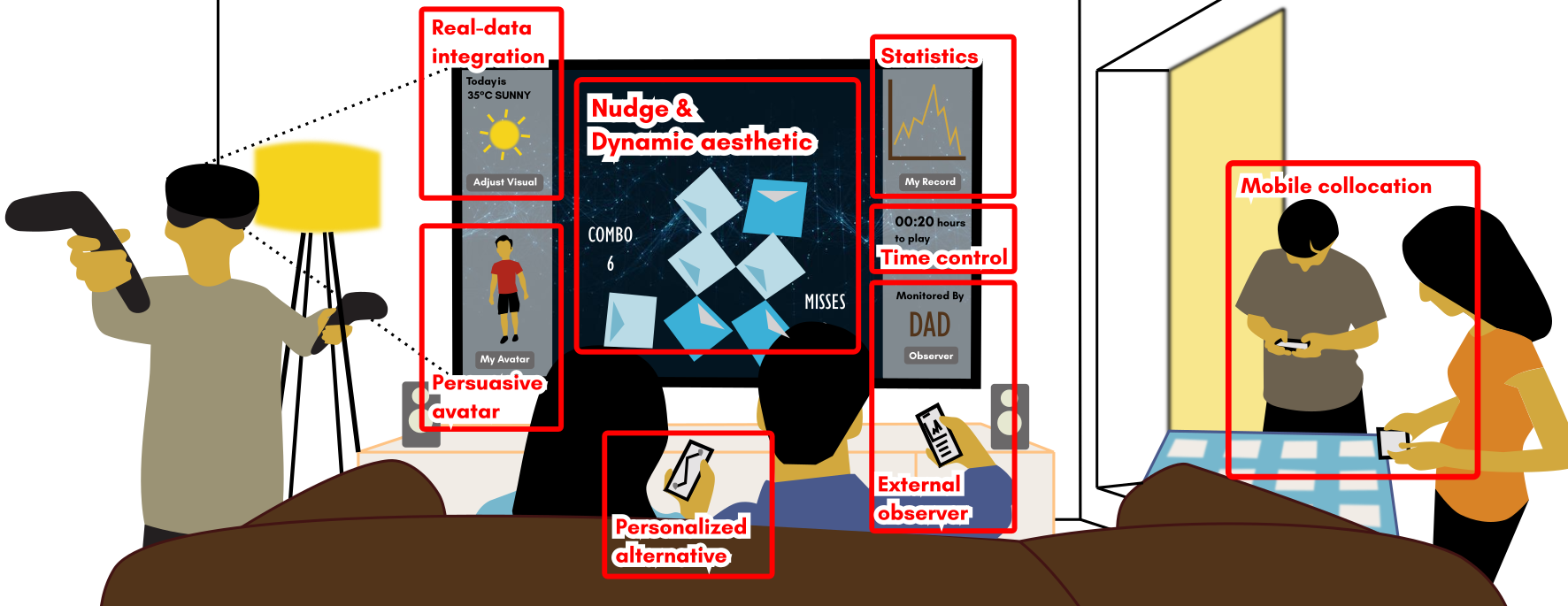}
  \caption{Persuasive elements of virtual user interface.}
  \label{fig:fig1}
}\end{figure}

The discussion of this review paper will start by elaborating on the concepts of technology addiction (\textbf{Section~\ref{sec:Addiction}}) and persuasive technology (\textbf{Section~\ref{sec:Persuasive}}). The latter will further scrutinize nine strategies for designing user interfaces (UIs) that can alleviate potential addiction (\textbf{Subsection~\ref{sec:Element}}). \textbf{Figure~\ref{fig:fig1}} illustrates the application of these strategies in the virtual experience. For example, a VR game may integrate nudges, avatars, and adaptive aesthetics. The gaming platform may also contain features such as players' statistics, time limit, and options to display real-world conditions. The statistics may also be shareable with external observers, such as family members. Likewise, AR applications may feature personalized alternatives and mobile collocation. This review paper collects the literature using PRISMA methodology~\citeAdd{page2021prisma} (\textbf{Section~\ref{sec:Methodology}}) and uses data visualization to justify the reason why such a topic was chosen (\textbf{Section~\ref{sec:Survey}}).

\section{Methodology}~\label{sec:Methodology}
This study employed the PRISMA methodology~\citeAdd{page2021prisma} to collect the articles. The collection was conducted in May 2022 from two credible databases for engineering research, the Association for Computing Machinery (ACM) digital library and the Institute of Electrical and Electronics Engineering (IEEE) Xplore digital. The literature collected contains papers with the following keywords in their titles:\newline

\noindent\textbf{[(Screen \textit{and} Time) \textit{or} Break \textit{or} Addiction \textit{or} Persuasive] \textit{and not} Drink \textit{and not} Heroin \textit{and not} Opioid \textit{and not} Drug \textit{and not} Alcohol}\newline

Despite being the scope of this review paper, the search keywords omit \textit{AR} and \textit{VR} words to reduce the inclusion of irrelevant results, for example, AR/VR literature that is unrelated to the persuasive technology. Moreover, the search keywords applied are adequate to include all literature whose title contains \textit{persuasive technology}, including its application in AR/VR.

\begin{figure}[t]
 \centering 
 \includegraphics[width=\columnwidth]{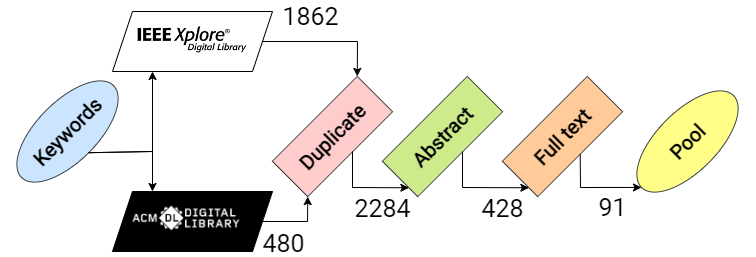}
 \caption{Selection diagram of PRISMA.}
 \label{fig:fig2}
\end{figure}

The articles collected are limited to only English literature from 2013 to 2022 to avoid the inclusion of obsolete studies. Literature collected from the ACM library are only \textit{conference papers}. On the other hand, literature from the IEEE library include; \textit{conferences}, \textit{journals}, \textit{early access}, \textit{articles}, \textit{books}, and \textit{magazines}. The following selection processes involve removing duplicates, examining abstracts, and reviewing the full texts. \textbf{Figure~\ref{fig:fig2}} shows the full selection flow, while \textbf{Table~\ref{tab:tab1}} explains the inclusion and exclusion criteria. The final process resulted in 91 articles.

\begin{table}[t]
  \caption{Inclusion and exclusion criteria.}
  \label{tab:tab1}
  \scriptsize%
	\centering%
  \begin{tabu}{%
	r%
	*{2}{c}%
	*{5}{r}%
	}
  \toprule
   \textbf{Inclusion} & \textbf{Exclusion} \\
  \midrule
    Addiction-related & Not a substance topic \\
    Ubiquitous technology & - \\
    Technology-related break taking	& - \\
    Technology-related screen time & - \\
    Screen-based persuasive content	& - \\
  \midrule
  \end{tabu}%
\end{table}

\section{Research Trend and Prior Surveys}~\label{sec:Survey}
This review study aims to encourage growth in the research about persuasive technology in AR or VR addiction. Currently, this topic suffers from being infrequently addressed as described in the following trends.

\begin{figure}[tb]
 \centering 
 \includegraphics[width=0.95\columnwidth]{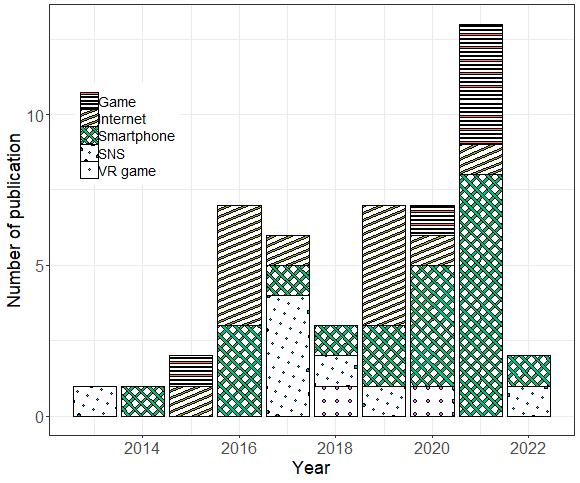}
 \caption{Topic distribution in technology addiction study.}
 \label{fig:fig3}
\end{figure}

\begin{figure}[tb]
 \centering 
 \includegraphics[width=0.95\columnwidth]{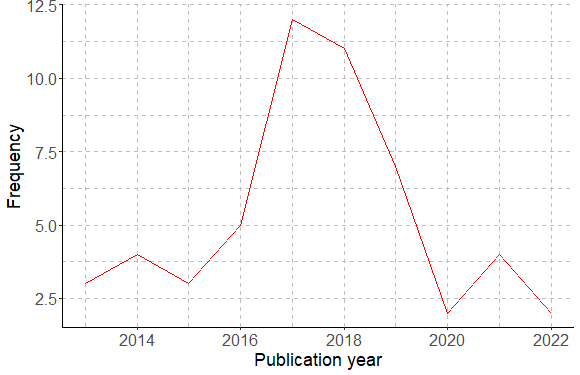}
 \caption{Research trend in persuasive technology.}
 \label{fig:fig4}
\end{figure}

\begin{figure}[tb]
 \centering 
 \includegraphics[width=0.75\columnwidth]{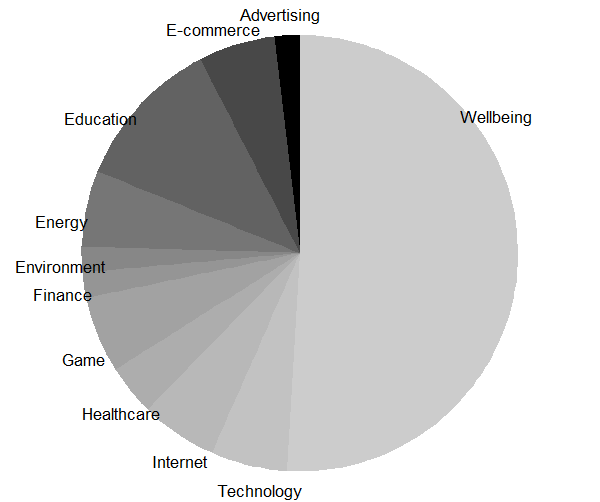}
 \caption{Topic distribution in persuasive technology.}
 \label{fig:fig5}
\end{figure}

\begin{itemize}
    \item \textbf{First}, the proportion of AR/VR topics in the study of technology addiction is low. \textbf{Figure~\ref{fig:fig3}} shows that articles on VR game addiction appear only twice in 2018 and 2020. The rest of the studies often focus on the addiction to the Internet, smartphones, and Social Networking Service (SNS). 
    \item \textbf{Second}, the trend of studies in persuasive technology experiences stagnant growth. \textbf{Figure~\ref{fig:fig4}} explains that persuasive technology was most studied in 2017 but has undergone a significant drop in 2020 that continues until today. 
    \item \textbf{Third}, topics about the game, the Internet, and technology are still minimum. \textbf{Figure~\ref{fig:fig5}} shows that, among the collected articles, well-being still occupies a substantial portion. 
\end{itemize}

The concept of persuasive technology in the case of technology addiction was also previously reviewed by Pinder \textit{et al.}~\cite{pinder2018digital} and Ndulue and Orji~\cite{ndulue2022games}. They similarly describe their work with the approach of \textit{behavior change}. They also elaborate on a comprehensive understanding of theory and its implication for technology design. Nevertheless, their work has yet to present the application of persuasive technology in AR/VR design. Thus, this opens more opportunities to explore persuasive technology for AR/VR applications, specifically, those that aim to reduce users' dependency.

\section{Understanding of Addiction}~\label{sec:Addiction}
Referring to Davazdahemami \textit{et al.}~\cite{davazdahemami2016addiction}, technology addiction is the change in human psychological state due to excessive dependency on technology reflected through the process of getting hooked, recovering, relapsing, and tolerating. Addiction is inherent to every technology user and varies by the level of severity. 

\begin{figure}[t]
 \centering 
 \includegraphics[width=0.7\columnwidth]{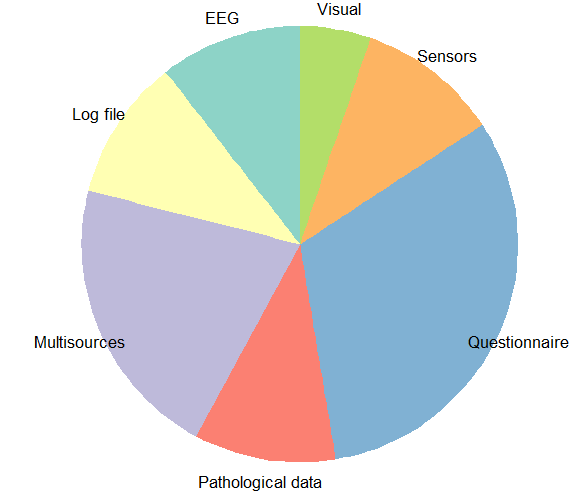}
 \caption{Proportion of detection methods for technology addiction.}
 \label{fig:fig6}
\end{figure}

\textbf{Figure~\ref{fig:fig6}} shows techniques to detect technology addiction. Questionnaires~\cite{inagaki2019preventive,lopz2020approaches,nawodya2022machine,sumalatha2016reducing,ulkhaq2018validity,valakunde2019prediction}, a common measurement tool in psychology, occupy about 32\% of the population or six out of nineteen. 
This proportion signifies technology addiction as a psychological ailment. Pozniak~\cite{pozniak2019real} also aligns with the finding that \textit{technology addiction is a behavioral addiction}. Moreover, researchers have been experimenting with emerging methods to measure technology addiction. Among them are the utilization of Electroencephalogram (EEG)~\cite{hafeez2017development,xiaoxi2021machine}, users' internet log data~\cite{purwandari2020internet,zhang2020correlation}, sensors~\cite{di2020multi,fletcher2016wearable,sen2016did,wibirama2017towards}, and the combination of them~\cite{ji2019real,krieter2022you,madushani2021screening,shae2020deep}. This implies that further integration of these sensors and intelligent logging devices can potentially understand the user status in the Metaverse. 

The nature of technology, which is ubiquitous~\cite{tiidenberg2017m}, rewarding and social~\cite{li2013they,luke2017exploring,serenko2020directing,urmanov2021empirical,vaghefi2014too,yang2020chinese}, and emitting \textit{flow}~\cite{soledad2020mediating} (the state of having undivided attention to the task on hand) makes it addictive. Thus, the level of recovery from addiction also depends on the level of dependency and users' awareness~\cite{qahri2021addiction}.

Curing technology addiction requires different approaches than the ones used for substance addiction because of their different nature. For example, the fourth row of \textbf{Table~\ref{tab:tab2}} shows that excessive use of substance leads to self-control impairment. On the other hand, addiction to technology leads to a lack of motivation. Users are less motivated to do productive tasks because they waste most of their time with technology. Thus, this shows that rehabilitation methods need features that motivate technology addicts to limit their consumption time. 

\begin{table}[t]
  \caption{Differences between technology and substance addiction~\cite{serenko2020directing}.}
  \label{tab:tab2}
  \scriptsize%
	\centering%
  \begin{tabu}{%
	r%
	*{3}{c}%
	*{5}{r}%
	}
  \toprule
   & \textbf{Substance} & \textbf{Technology} \\
  \midrule
    \textbf{Regulation} & Largely regulated & Less restricted \\
    \textbf{Risk} & Lethal health risk & Multi-dimensional risks \\
    \textbf{Avoidance} & Easy & Difficult \\
    \textbf{Implication} & Self-control impairment & Lack of motivation \\
    \textbf{Effect} & All ages & Juvenile and youth \\
  \midrule
  \end{tabu}%
\end{table}

\section{Persuasion in Virtual Technology}~\label{sec:Persuasive}
Persuasive technology is a combination of persuasion and technology~\cite{murillo2018framework}. The inherent persuasion characteristic of technology makes it naturally addictive~\cite{murillo2018framework}. Thus, with the same attribute, the designer can make technology positively persuading to alleviate the addiction.

\subsection{Concept of Persuasive Technology}~\label{sec:Concept}
The theory of persuasive technology originates from three basic principles~\cite{simons2016psychological}:
\begin{itemize}
    \item\textbf{Planned Behavior:} The main objective of persuasion is to change the routine or behavior of the users. To succeed, users need to have the intention to change triggered by society.
    \item\textbf{Self Determination:} Users need autonomy and competence to realize their goals. Their methods also need to be correlated with the ambient situation. When fulfilled, users can conduct the change himself. Else, users need external enforcement.
    \item\textbf{Control:} Users control their change progresses by comparing the ambient situation with their current states. They improve and develop the behavior by filling in the gap.
\end{itemize}
 
Persuasive technology is the combination of two disciplines, the human factor, and psychology. Nevertheless, there is a discrepancy between the main idea of persuasive technology and the human factor. The objective of the human factor is to develop technology that fits human needs. On the other hand, persuasive technology is the technology that influences human transformation to suit the technology. The web of persuasive technology involves aspects of individual factors, beliefs, outcome expectations, and social factors reflected in the user interfaces (UIs)~\cite{torkamaan2021integrating}.

\begin{table}[t]
  \caption{Persuasion strategies~\cite{emets2017using}.}
  \label{tab:tab3}
  \scriptsize%
	\centering%
  \begin{tabu}{%
	r%
	*{2}{c}%
	*{8}{r}%
	}
  \toprule
    \textbf{Strategies} & \textbf{Description} \\
  \midrule
    \textbf{Reduction} & Simplification of process step \\
    \textbf{Tunneling} & Step-by-step guidance on a task \\
    \textbf{Tailoring} & Content adaptation to user’s interest \\
    \textbf{Suggestion} & Information of the right condition to do the task \\
    \textbf{Self-monitoring} & Limitation of usage time through mindful reflection \\
    \textbf{Surveillance} & Systems' observation of user’s behavior \\
    \textbf{Conditioning} & Reward or punishment \\
  \midrule
  \end{tabu}%
\end{table}

\textbf{Table~\ref{tab:tab3}} explains strategies done by the UIs that trigger human transformation~\cite{emets2017using}. These range from the subtle approach such as \textit{tailoring} to obtrusive ones such as \textit{surveillance}. To function accordingly, the design of UI requires the embodiment of persuasion attributes that are social~\cite{fritz2014persuasive,drossos2019online,van2020persuasive,orji2019drivers}, flexible~\cite{van2021experiences}, gamified~\cite{van2021experiences,ndulue2021std,nystrom2017gamification,gu2015application}, autonomous~\cite{haque2016persuasive}, demonstrative~\cite{shih2017selecting}, remindful~\cite{shih2017selecting}, verifiable~\cite{ndulue2021gender}, concise~\cite{mongadi2022persuasive}, relevant~\cite{mongadi2022persuasive,haller2017energy}, incremental~\cite{murillo2018framework}, subtle~\cite{murillo2018framework}, and virtual~\cite{sara2019study}. These attributes show that persuasive technology is dynamic and customized. Therefore, it is \textit{one-for-one} technology instead of a \textit{one-for-all}. Furthermore, persuasive technology also needs to account for the users' demographic factors such as genders~\cite{orji2019drivers,muhammad2018personalizing,oyibo2018perceived}, ages~\cite{muhammad2018personalizing}, personalities~\cite{ciocarlan2017qualitative,orji2017towards}, cultures~\cite{oyibo2017influence,hong2017improved}, ethnicities~\cite{oyibo2018susceptibility}, and state of changes~\cite{oyebode2021tailoring}.

\subsection{Persuasive Elements in UI Design}~\label{sec:Element}
Alhammad and Gulliver~\cite{alhammad2013context} proposed multi-layered persuasive design elements of relevance, personalization, and customization. \textit{Relevancy}, the innermost layer, acts as a hook for users' trust. It easily influences the users if the UIs blend well with their activities. The \textit{personalization} in the middle layer uses UIs to comfort the users and engage them by behaving in a way that fits with users' identities. \textit{Customization} in the outermost layer aims to build users' habits by connecting the individual to societal factors. These mechanisms present the following UI design components.

\subsubsection{Nudge}~\label{sec:Nudge}
Nudge is an unobtrusive visual cue in the UIs designed to direct users to perform a particular task to get the positive outcome desired by the designer. A persuasive nudge immerses with the circumstance, has a clear target and appearance, progresses from obscure to obvious, and employs multiple feedback channels. For example, in work by Schneider and Graham~\cite{schneider2015pushing}, nudge appears to motivate the players of static bike exergame (gamified physical exercise) to regulate their heartbeats. Therefore, nudge appears as a point-collecting rainbow positioned at the lower part of the game screen to influence the players' avatars to go lower by reducing their cycling speed.\newline

\begin{figure}[t]
 \centering 
 \includegraphics[width=0.75\columnwidth]{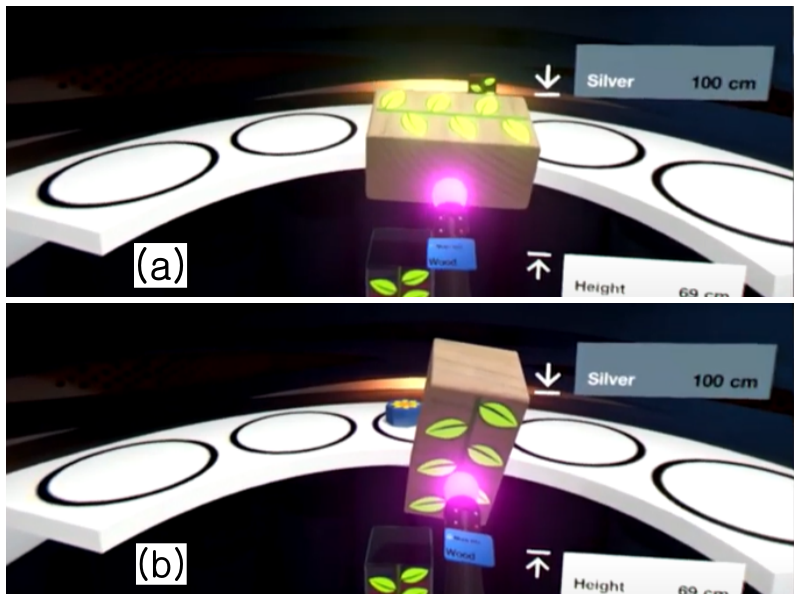}
 \caption{Nudge elements in Tumble VR~\protect\citeAdd{playstationaccess_2016}.}
 \label{fig:fig7}
\end{figure}

\noindent\textbf{Strategy and example in the virtual environment:}
AR/VR game developers may employ the technique of integrating a nudge that fuses into the game setting. The nudge may take the form of the typical game component, for example, a part of the main character. It is preferable that nudge only appears for a few moments throughout the game session, such as at a higher level or after playing for a few hours. Nudge that shows too often is obtrusive and will not fit into the game scenario. This technique can subtly direct the players to choose to break from the game. Developers also need to consider that nudge leads to an open-ended act that provides flexible choices to the players. For example, the players gain a score if they follow the nudge but lose nothing if they leave it. It differs from the gamification technique that forces the players to act according to the nudge to score. Many AR/VR game developers today, for example Pokémon Go, Beat Saber, and Tumble VR, insert a nudge in their applications~\citeAdd{meschtscherjakov2017pokemon,schneider2017nudging}. \textbf{Figure~\ref{fig:fig7}} shows an example of a nudge in Tumble VR\footnote{\url{https://store.playstation.com/en-hk/product/HP9000-CUSA05314_00-TUMBLEVR00000001}}. The rectangular block persuades the players to stack in the vertical position to increase the height. However, if the lower foundation is loose, this positioning may cause the stack to tumble. The players can choose whether to stack vertically and lose the game or pile horizontally but produce a lower tower height.

\subsubsection{Users' statistics}~\label{sec:Statistic}
The presence of users' statistics in the UI design aims to raise users' awareness of their ability to self-monitor their technology usage rates. \textit{Time} is the most common statistic implemented in the UI. It informs the users about how many hours they have consumed using the technology. Adib \textit{et al.}~\cite{adib2021persuasive} also included other statistics for gaming platforms, such as remaining time to use per game, players' classification of addiction, total merchandise spending rate, discussions about self-monitoring through community sharing, and suggestions from the experts.\newline

\begin{figure}[t]
 \centering 
 \includegraphics[width=0.75\columnwidth]{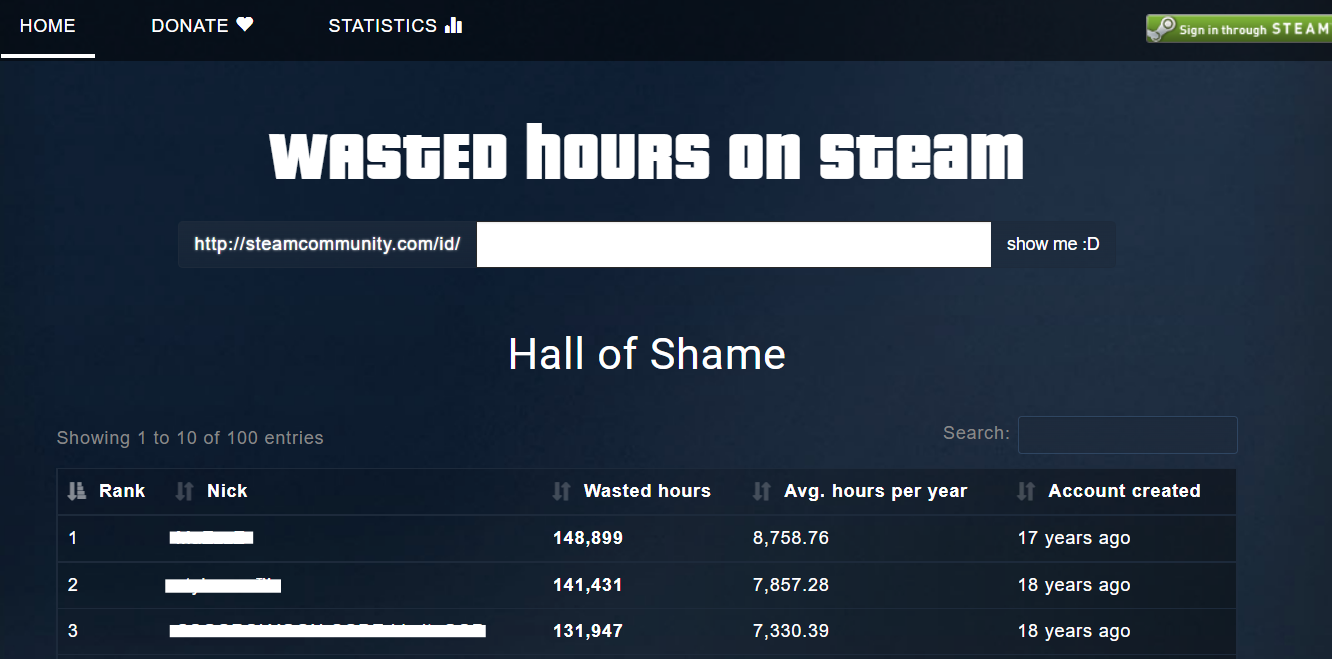}
 \caption{SteamTime gameplay statistics~\protect\citeAdd{steamtime}.}
 \label{fig:fig8}
\end{figure}

\noindent\textbf{Strategy and example in the virtual environment:}
Developers of AR/VR gaming platforms or manufacturers of a head-mounted display (HMD) may record gameplay statistics by utilizing the sensor in the HMD and the game log data. They can display this data on the users’ profiles in the form of total gameplay hours and records for each game. Not only this will help users to acknowledge their gameplay addiction but also assess in which game they should reduce their playtime. Developers may also include social features such as the rank of gameplay records among the users’ circle and forums where users can share their stories of overcoming addiction. \textbf{Figure~\ref{fig:fig8}} shows an example of this feature in SteamTime, the third-party application retrieving gameplay data from Steam. It displays users' gameplay records and game ranks. Game players commonly use this to measure their level of gaming addiction because they can see their records of hours and dollars.

\subsubsection{Mobile collocation}~\label{sec:Collocation}
A mobile collocated application is an application that can only display a section of visual content on a smartphone. To enjoy the full content, the user needs to group with others by physically pairing their smartphones. This technique promotes users' socialization while keeping them engaged with the application content. Lucero \textit{et al.}~\cite{lucero2015mobile} showed the implementation of this feature in photo viewing application and mobile gaming. Their work displays how a landscape photo splits into four. The complete picture is visible when four users collocate their smartphones together in the same place.\newline

\begin{figure}[t]
 \centering 
 \includegraphics[width=0.75\columnwidth]{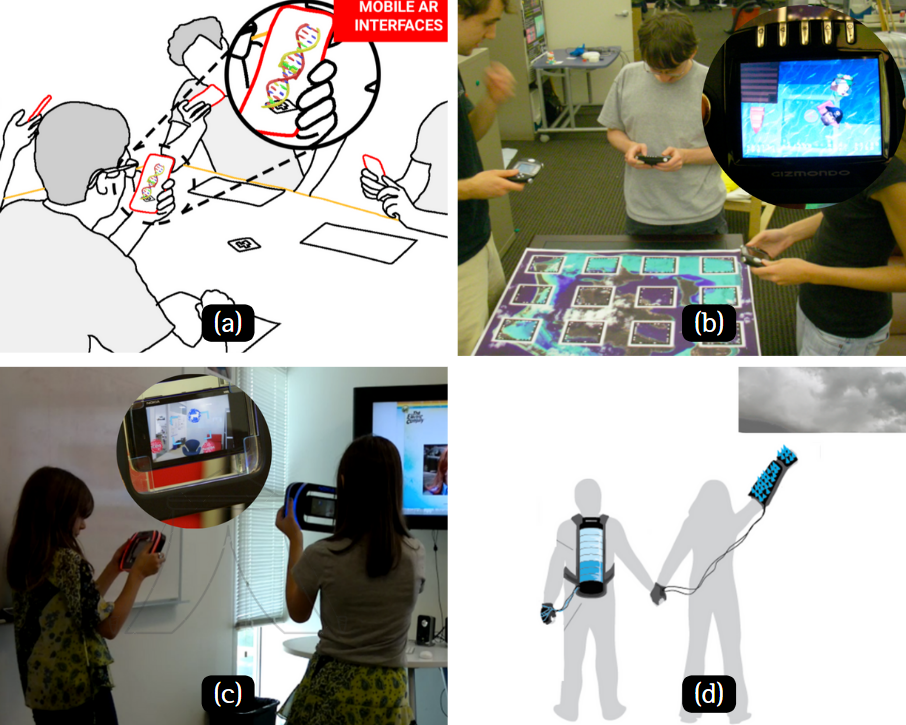}
 \caption{Mobile collocated application; (a) CollabAR~\protect\citeAdd{wells2020collabar}, (b) BragFish~\protect\citeAdd{xu2008bragfish}, (c) Electric Agents~\protect\citeAdd{ballagas2013electric}, and (d) Hotaru~\protect\citeAdd{isbister2017interdependent}.}
 \label{fig:fig9}
\end{figure}

\noindent\textbf{Strategy and example in the virtual environment:}
AR developers may create an application that splits virtual characters among different users' screens. Therefore, users need to collaborate to find the character. This technique is applicable in many applications. Game developers may create a pair of game characters, put them on separate users' screens, and write a scenario where these two characters must act together to score. Additionally, exhibition developers may split one virtual visual into different screens. Users then can team up to enjoy the full content. \textbf{Figure~\ref{fig:fig9}} shows examples of collocation-based AR/VR in an application named \textit{BragFish}. It integrates a collocation function for a mobile AR board game~\citeAdd{xu2008bragfish}. A group of players move their smartphones above the board to row the virtual boat and find the fish. Since the fishes constantly move, the group needs good teamwork. The group scores by capturing as many fish in a given time. There are also other examples of collocation-based AR/VR such as \textit{CollabAR} (mobile collocation system for the 3D-model group viewing activity)~\citeAdd{wells2020collabar}, \textit{Electric Agents} (a mobile AR interaction for educational television (TV) programs)~\citeAdd{ballagas2013electric}, and \textit{Hotaru} (the wearable energy collecting game that collocate by pairing the suit with a group of two players)~\citeAdd{isbister2017interdependent}

\subsubsection{Personalized alternatives}~\label{sec:Alternative}
When nudges and collocations are infeasible, developers can replace persuasive elements with personalized alternatives. For example, Gupta \textit{et al.}~\cite{gupta2021persuasive} proposed a mobile navigation service that integrates log data and preferences. While persuading them to reduce their carbon footprint, the service only suggests familiar choices. For example, transport alternative for the users used in driving places driving choice on top while also informing its the potential carbon emission. This approach not only persuades users to take greener transportation modes but also educates them about their habits. Bothos \textit{et al.}~\cite{bothos2016recommender} developed a similar application. For example, their navigation service only suggests the users go on foot if the distance is within the users' historical record of walking activity.\newline

\begin{figure}[t]
 \centering 
 \includegraphics[width=0.75\columnwidth]{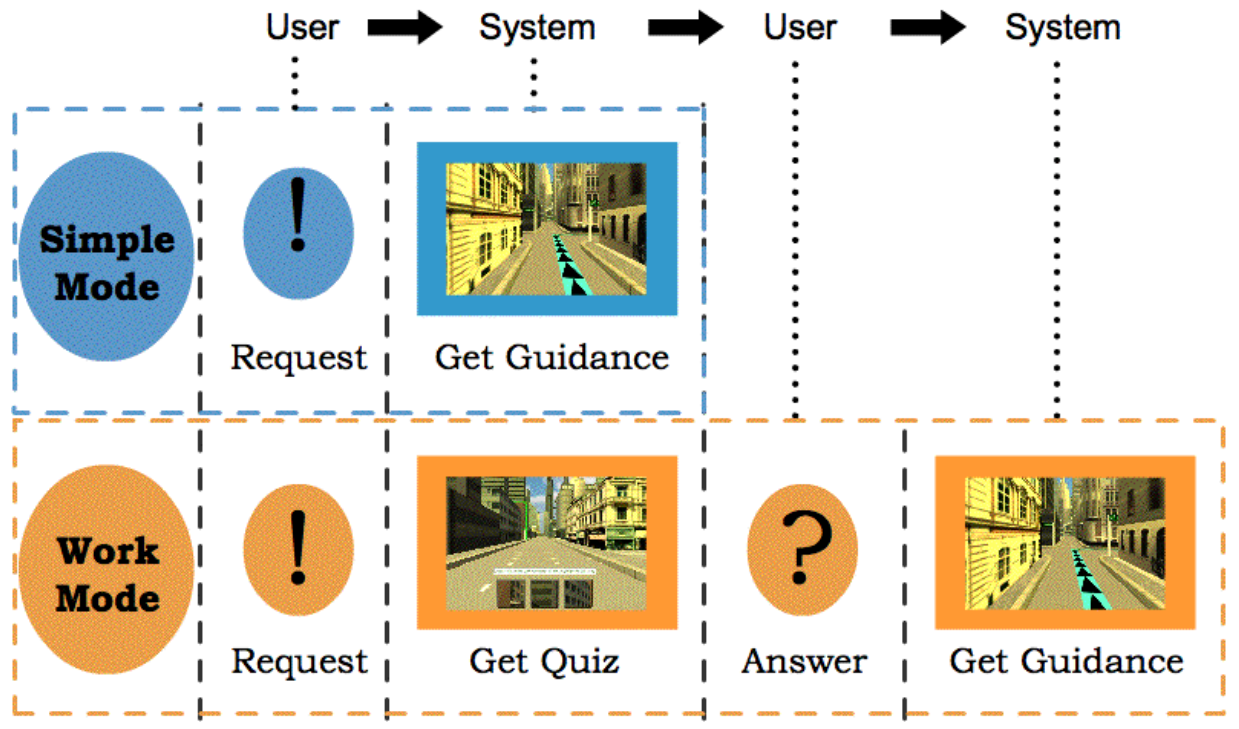}
 \caption{Personalized alternative navigation system~\protect\citeAdd{wen2014really}.}
 \label{fig:fig10}
\end{figure}

\noindent\textbf{Strategy and example in the virtual environment:}
AR developers may implement this technique for a navigation system. The application can display two navigation options, passive guidance, and active search. Users who have trouble recognizing their surroundings may choose the first option. The second one targets users who are unfamiliar with map reading. In the second option, developers may encourage users to create their AR route by only hinting to them with the picture of nearby landmarks to follow. Aside from engaging users through gamification, this methodology also helps with human spatial memory~\citeAdd{wen2014fighting}. Wen \textit{et al.}~\citeAdd{wen2014really} proposed a hard-to-use AR-based navigation system to lessen users' dependency on navigation tools. \textbf{Figure~\ref{fig:fig10}} illustrates that users can choose to navigate using one of the two alternatives, simple and work. If users pick a simple alternative, an AR guide will show on the phone screen to help with the navigation. Nevertheless, users need to solve a quiz about identifying nearby landmarks before the AR guide appears if they select a work alternative.

\subsubsection{External observer}~\label{sec:Observer}
When the persuasive relationship between users and technology fails, an external observer is required to trigger enforcement for the users. However, it is preferable that the observer is someone close to the users and can have physical communication, for instance, with a family member. Katule \textit{et al.}~\cite{katule2016leveraging} demonstrated a \textit{social rapport} feature in their wellbeing application where both the user and observer use the same application. Toward the users, the application only gathers their pathological data and suggests methods to maintain a healthy lifestyle. While toward the observers, the application sends reminders and notification about user's performance.\newline

\begin{figure}[t]
 \centering 
 \includegraphics[width=0.6\columnwidth]{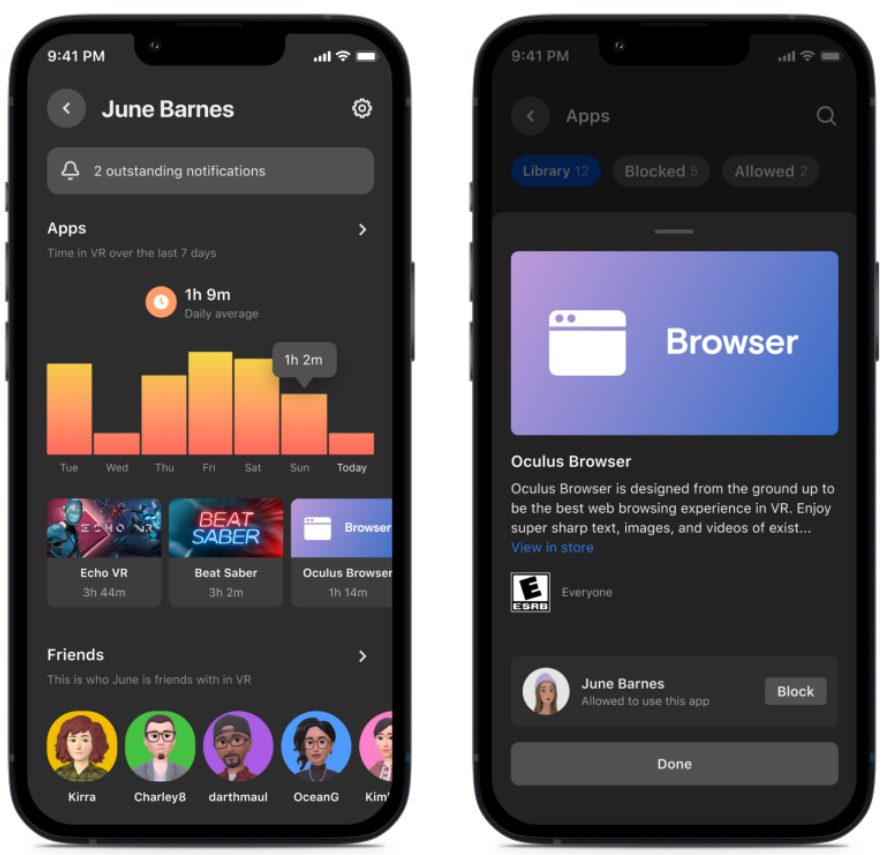}
 \caption{Parental supervision dashboard on Meta Quest~\protect\citeAdd{meta_2022}.}
 \label{fig:fig11}
\end{figure}

\noindent\textbf{Strategy and example in the virtual environment:}
This technique can be paired with the players' statistics. It is commonly known as the parental monitoring feature. Developers of AR/VR gaming platforms may provide a monitoring account feature through which the players' family members can sign up. They can monitor and control the virtual activities of the people they observe on the platform or using a standalone mobile application. Not only will they get notifications about the gameplay records, but also they can set the gameplay limit as desired. For example, Meta introduced a parental supervision tool in March 2022 on Meta Quest, formerly known as Oculus~\citeAdd{meta_2022}. \textbf{Figure~\ref{fig:fig11}} shows the dashboard of Meta Quest’s gameplay statistics. This feature monitors the total time teens spend on the VR game by collecting data from the Quest headset. It also displays the daily time spent on VR and the total hours spent on each VR game on the dashboard.

\subsubsection{Adaptive aesthetics}~\label{sec:Aesthetic}
Persuasive aesthetics adaptively change following the state of the users, hence called \textit{adaptive aesthetic}. Han \textit{et al.}~\cite{han2016pleasurable} present this example in an exergame. The monitor screen on top of the static bike shows different scenery depending on the users' heart rates. A flowery scene that appears under a normal heart rate will change into a barren landscape as it approaches the anaerobic stage. This change aims to trigger motion sickness and persuade the users to reduce their speeds or stop the games. Oyebode \textit{et al.}~\cite{oyebode2020nourish} also use a tree that grows denser as the players increase their daily steps. In the energy field, Liu~\cite{liu2013bluepot} and Vilarinho \textit{et al.}~\cite{vilarinho2016combining} created a flower that withers following the rise in domestic energy consumption. In financial applications, Chaudhry and Kulkarni~\cite{chaudhry2022robinhood} use a forest that becomes denser when the users increase their investment rates.\newline

\begin{figure}[t]
 \centering 
 \includegraphics[width=0.75\columnwidth]{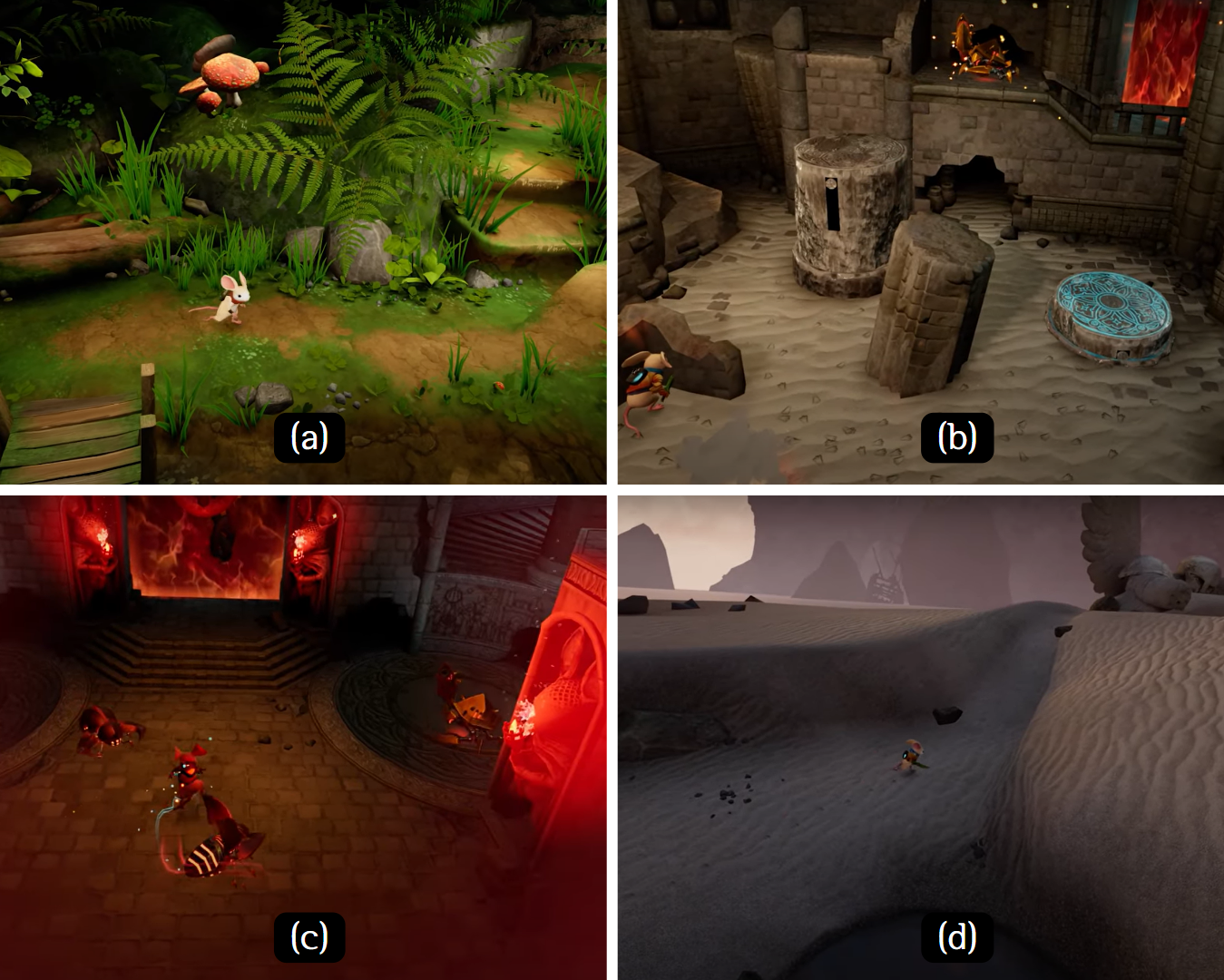}
 \caption{Adaptive aesthetics in Moss; (a) Start of the game, (b) after 30 minutes, (c) after 45 minutes, and (d) after 1 hour and 15 minutes~\protect\citeAdd{99th_vr_2021}.}
 \label{fig:fig12}
\end{figure}

\noindent\textbf{Strategy and example in the virtual environment:}
Implementation of adaptive aesthetics may be intricate since different people view aesthetics differently. Thus, developers need to refer to the principal game concept, whether it is more of a beautiful game or a disturbing one. Developers of AR/VR beautiful games may consider gradually changing the scenery to a dull one. On the other hand, developers of games whose concept is violence may replace the visual with cutesy. Either way, developers need to implement incremental change based on the time players spend in the virtual game. For example, Moss, an action venture puzzle VR-based game, persuades the players to break the \textit{flow} using adaptive aesthetics by changing its interface from the beautiful scenery to the tense one. \textbf{Figure~\ref{fig:fig12}} shows that the game starts in a serene green forest scenery. As the player spends more time in the game, the aesthetics gradually change from the beautiful forest to the arid desert.

\subsubsection{Persuasive avatar}~\label{sec:Avatar}
An avatar makes the system's interaction with the users more humane. A visual design that contains avatars is more appealing than one that 
does not~\cite{ibrahim2013dual, xiao2019should, aydin2018couch}. \textbf{Table~\ref{tab:tab4}} outlines specific features of persuasive avatars. In general, they are assessed by how they sound and move. For example, the avatar will normally speak slowly in a low-pitched voice. They will also mimic the human current state or behave in a way that human needs to follow. \newline

\noindent\textbf{Strategy and example in the virtual environment:}
To break the \textit{flow}, VR developer needs to use the inversion of the feature mentioned in \textbf{Table~\ref{tab:tab4}} to create a perception of an unappealing avatar. They may insert this technique at several parts of the game, for example, after playing for hours. The rest of the game may have the usual avatars. It aims to make the player uncomfortable and finally break the flow. An example of this technique in a VR-based game is \textit{Weeping Doll}, a story-driven horror adventure game. The game received a low rating due to the unpleasant acting of its two avatars~\citeAdd{darkzero_2016}. 
The ghost maid avatar whose role is to guide the players with navigation behaves irritatingly by repeatedly telling the players where to go. Another avatar, a child ghost, produces a voice unfit for the game ambiance. Its high-pitched voice fails to create a spooky feeling.

\begin{table}[t]
  \caption{Features of persuasive avatar.}
  \label{tab:tab4}
  \scriptsize%
	\centering%
  \begin{tabu}{%
	r%
	*{2}{c}%
	*{12}{r}%
	}
  \toprule
    \textbf{Features} & \textbf{Properties} \\
  \midrule
    \textbf{Speech} & Follows users' dialect~\cite{khataei2015personalized}, slow~\cite{dubiel2020persuasive} \\
    \textbf{Tone} & Low pitch~\cite{wang2018effect}, low volume~\cite{wang2018effect}, low depth~\cite{dubiel2020persuasive} \\
    \textbf{Voice source} & Recorded~\cite{parmar2020navigating} \\
    \textbf{Conversation} & Personal, affective, and logical~\cite{shi2020effects,saunderson2019would} \\
    \textbf{Gesture} & Follows emotion, peer-like~\cite{xiao2019should,kantharaju2018two}, mimics users~\cite{jain2020kyro} \\
    \textbf{Impression} & Extrovert~\cite{hyde2015using}, follows culture~\cite{zhou2017adapting} \\
  \midrule
  \end{tabu}%
\end{table}

\subsubsection{Real-world data integration}~\label{sec:Data}
Presenting real-world situation in the virtual environment aims to inform the users about their surroundings and influence them to decide accordingly. For instance, Coulton \textit{et al.}~\cite{coulton2014designing} integrated real-climate data into an environmental protection game. The game visual changes to follow 
the actual climate data. It shows that real-world data integration into a game can control the players' attitudes to the game content and player's engagements to the game \textit{flow}.\newline

\begin{figure}[t]
 \centering 
 \includegraphics[width=0.75\columnwidth]{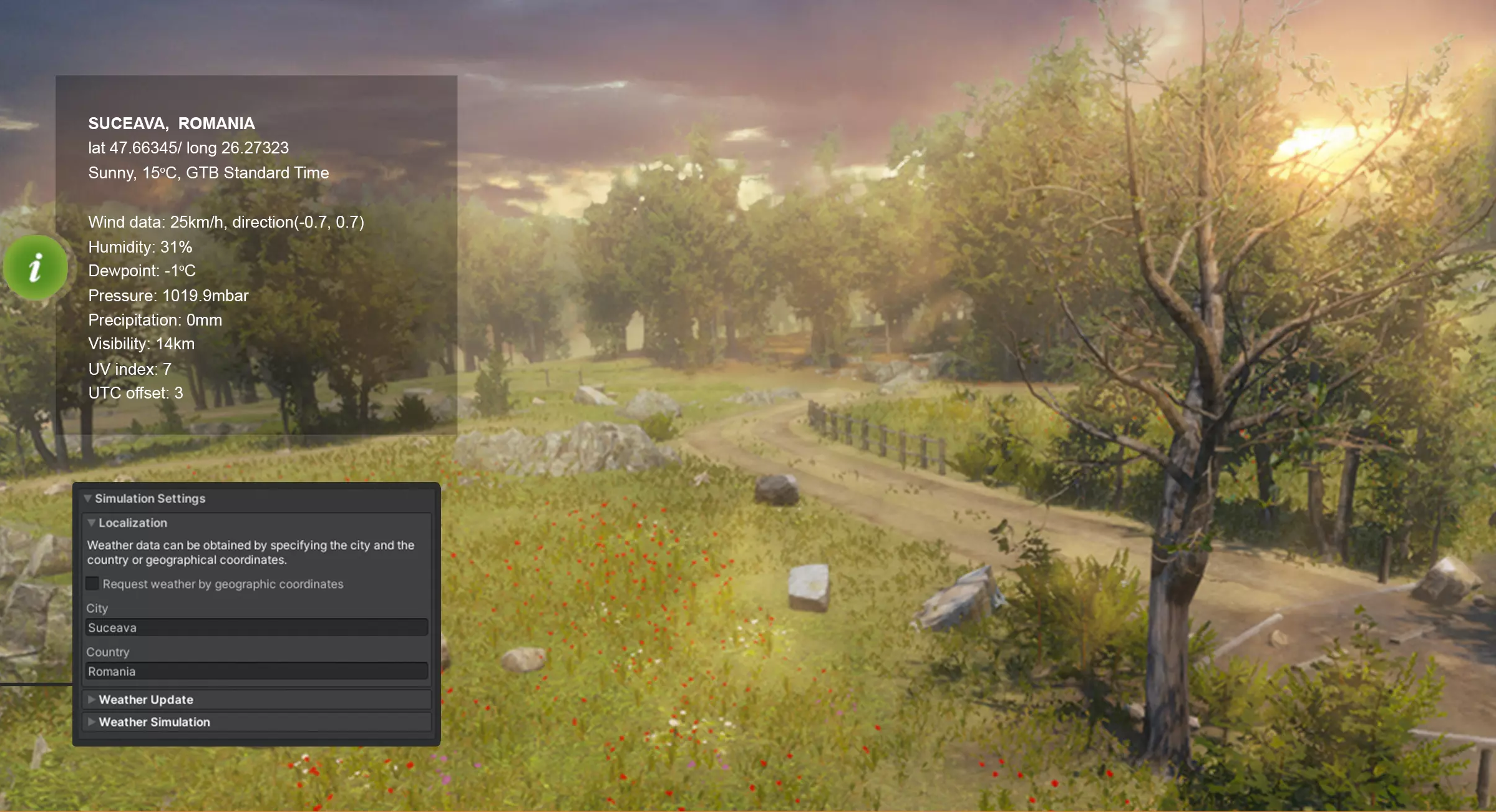}
 \caption{Interface of \textit{Real-Time Weather} plugin~\protect\citeAdd{zambalic_2021}.}
 \label{fig:fig13}
\end{figure}

\noindent\textbf{Strategy and example in the virtual environment:}
AR/VR developers may utilize actual data plugins in the virtual development platform store. They can choose which data to integrate, from which location, and into which part of the game. Developers may also add options in the application whether the players want to see the visual interface that integrates with actual data or not. From the developer's side, these plugins can save development time. Users can get information about their surroundings while being in the virtual environment. The technique can also encourage them to break from the virtual world. \textbf{Figure~\ref{fig:fig13}} gives an example of \textit{Real-Time Weather}, a plugin from Unity that enables VR developers to create a visual that feels akin to the actual weather condition of a particular city.

\subsubsection{Time limit}~\label{sec:Time}
Time limit is a common technique employed in many internet applications. This technique works with users setting the time limit of how long they can use a particular application. Users will need to quit the application when the limit is exceeded. This technique is also combinable with other features. For example, using human pathological data to determine limits~\cite{fukushima2021break}, implementing both hard and soft time limit ~\cite{chisler2016handle,tseng2019overcoming}, allowing users to negotiate the limit~\cite{luo2018time}, providing in-application break activity instead of quitting~\cite{thai2020use}, and using robots to notify the limit~\cite{abdi2022gamer,zhang2020socially,kucharski2016apeow}.\newline

\begin{figure}[t]
 \centering 
 \includegraphics[width=0.75\columnwidth]{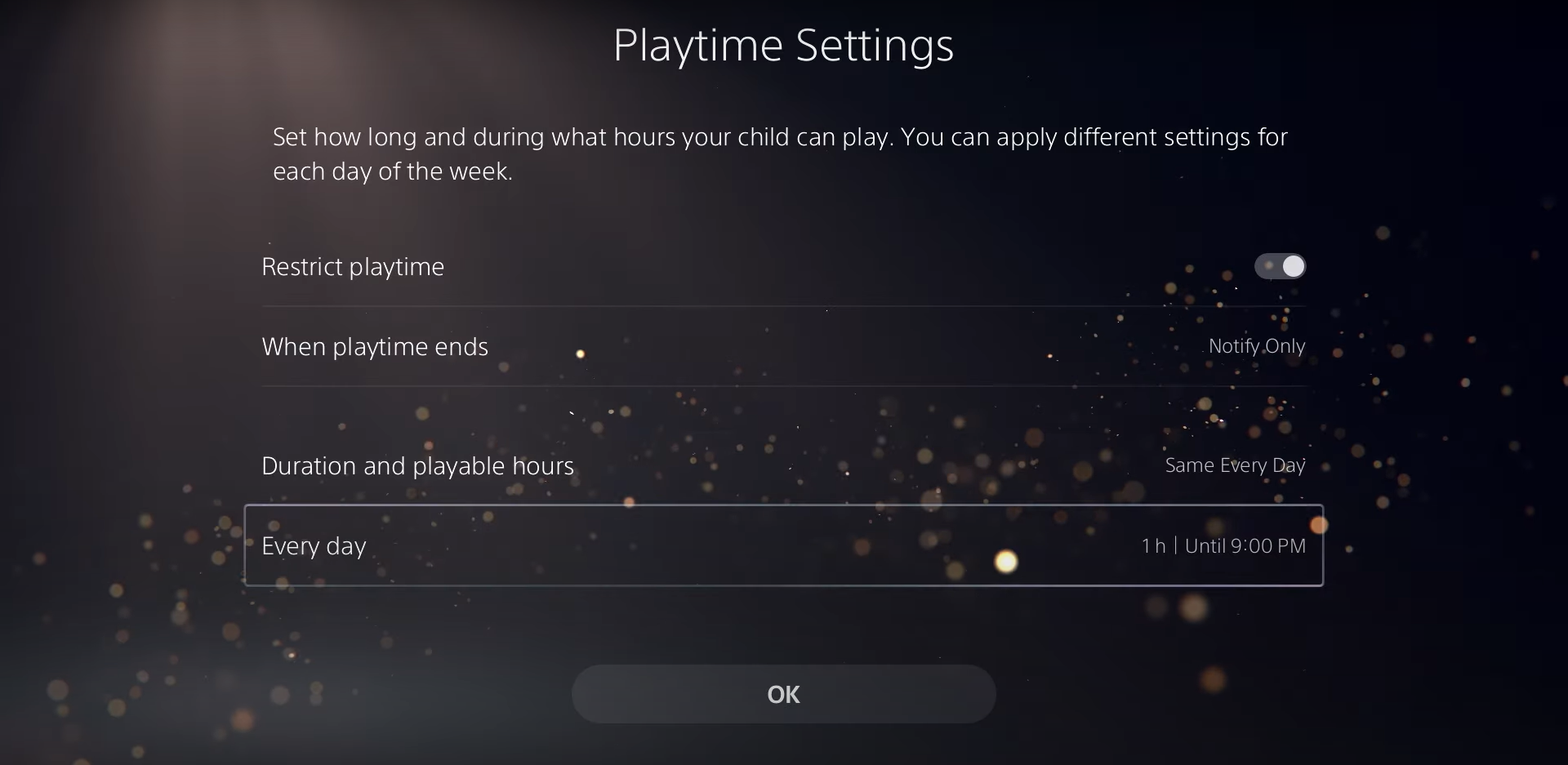}
 \caption{PlayStation5 console's parental control~\protect\citeAdd{playstation_2021}.}
 \label{fig:fig14}
\end{figure}

\noindent\textbf{Strategy and example in the virtual environment:}
AR/VR developers may implement this technique as an extension of the external observer feature. They may improve the methodology to set the time limit by measuring the users' actual state instead of only setting the hourly limit. This function can utilize the data recorded by the HMD or mobile phone sensors in real-time. Alternatively, this data can inform the average number of hours users can safely spend in the virtual environment. Later, observers may set the time limit based on this reference. \textbf{Figure~\ref{fig:fig14}} shows an example of PlayStation's parental control section providing a time limit feature where parents can set the allowable time for their children to use the console. The sensor in the VR headset records the time users spend using it and integrates it with the time limit being defined.

\section{Conclusion}~\label{sec:Conclusion}
The existing literature on persuasive technology supports the technique of unobtrusively influencing humans to perform positive actions. Although today they are common in well-being, it has prospective to integration to the topic of addiction in the virtual environment. The development of nine persuasive elements in the UI design discussed here also provides a new perspective on dealing with dependency in AR/VR applications. Further, developers may use these elements as guidelines to design virtual applications that are not only engaging but also less addictive.



\bibliographystyle{abbrv-doi-hyperref-narrow}

\bibliography{bibliography}

\bibliographystyleAdd{abbrv-doi-hyperref-narrow}
\bibliographyAdd{addbib}

\end{document}